\documentstyle[preprint,aps,axodraw]{revtex}
\tightenlines

\begin{document}

\draft
\title{On nonleptonic decays of Supermultiplets}
\author{R.~Delbourgo$^*$ and Dongsheng Liu$^{**}$}
\address{School of Mathematics and Physics, University of Tasmania,\\
 Hobart, Australia 7001}
\date{\today }
\maketitle

\begin{abstract}
By describing strong interactions between hadrons via a relativistic 
supermultiplet scheme and regarding weak interactions as a perturbation thereof,
we derive expressions for nonleptonic weak decay amplitudes in terms of 
constituent quark masses and CKM angles, with no other parameters. Application 
of this method leads to $\Delta I=1/2$ dominance in some 
pseudoscalar meson decays if one scales down the couplings of heavy particles 
by $\sqrt{M}$ mass factors, in keeping with heavy quark theory expectations.
However, certain B and D decay processes to kaons are badly predicted and
point to substantial soft gluon renormalization effects in W-quark 
interactions.
\end{abstract}


\narrowtext

\section{Introduction}  

With so much data available for the decays of heavy quark composites, a real 
industry has grown, devoted to calculating nonleptonic weak amplitudes. A 
standard picture has evolved (Neubert 1994, 1997; Neubert and Stech 1997;
Stech 1997) from the work of Bauer, Stech and Wirbel (1987), which parametrises
flavour-changing decays in terms of a number of effective four-quark operators,
such as
$$ {\cal H} = \frac{G_F}{2\sqrt{2}}V_{UD}V^*_{ud}[c_1(\mu)(\bar{d}\gamma u)_L.
 (\bar{U}\gamma D)_L + c_2(\mu)(\bar{U}\gamma u)_L.(\bar{d}\gamma D)_L],$$
and associated parameters $c_i(\mu)$; the idea is then to connect the running
$c(\mu)$ values via QCD and the renormalization group with heavy quark theory 
(Isgur and Wise 1989, 1990) schemes. This description (Ali et al 1998, 1999;
Lu 1999) introduces many parameters and requires the input of meson decay 
constants $f$ (which are extracted from experiment). Even after introducing
so many phenomenological constants, a certain amount of confusion remains in 
this subject, mainly connected with the role of non-factorizable contributions 
(Beneke 2000, Neubert 2000); indeed some of the elaborations are so Byzantine 
in their intricacies that one yearns for a simpler and more direct approach
to this problem. The only striking and mysterious fact remains 
the dominance of the $\Delta I=1/2$ rule in kaon and hyperon decays.

In this paper we will attempt to shed some light on this topic by trying a
different approach, one which has long been applied to electromagnetic 
interactions; namely we will regard the Lagrangian of electroweak theory as 
a perturbation of the strong interactions. The difficulty with such an approach
lies in knowing the values of strong interaction amplitudes, since this is to 
be the starting point for applying perturbation theory. Strong amplitudes are 
of course given reliably by QCD at high energy, but at low energy we are 
obliged to resort to models which respect the symmetries of QCD, rather than 
QCD itself, because of unknown effects produced by colour confinement and
soft multigluon exchanges that enhance the quark interactions at low mass 
scales. One of the simplest such models is the linear (or nonlinear) sigma-type
model (Scadron 1997) because it embodies full chiral symmetry for massless bare
quarks.  Another favourite scheme is the spin-flavour supermultiplet model, 
which not only applies to heavy quarks but to light ones as well (Salam et al 
1965; Sakita and Wali 1965), provided their masses are dressed to their 
constituent values. We explore this route and take it that strong amplitudes 
are quite well described---often to better than 10\%---by supermultiplet tree 
interactions (Delbourgo and Liu 1996); then we shall add electroweak 
interactions as a small perturbation\footnote{A preliminary version of this 
method was presented in Delbourgo and Liu (1998), but publication requirements 
meant that the articles were so compressed as to be difficult to follow; the 
details are fleshed out here and the scope greatly expanded.}.

Semileptonic decays are well-understood via weak-boson exchange, provided
the hadronic matrix element $\langle f|J^{weak}|i\rangle$ of the weak current 
$J^{weak}$ is extracted from experiment or estimated reliably from theory, 
using heavy quark symmetry, dispersion relations, sum rules, $1/N_c$ expansions 
or whatever other tool one can make use of. We will have nothing to say about 
semileptonic processes; rather, our focus is on nonleptonic processes where 
weak and strong interactions are linked. Our attitude towards flavour changing 
weak amplitudes is that, to first order in the Fermi coupling $G_F$, the 
process can be construed as a set of W-exchange loop diagrams between the 
hadronic participants, with the {\em hadronic} amplitude satisfactorily 
determined by higher spin-flavour symmetry interactions. The computational 
rules are thus fixed and our answers can only depend on the masses of the 
constituent quarks and the CKM mixing angles. This assumes the Feynman 
integrals are convergent---which they are, thanks to the unitarity of the CKM 
matrix\footnote{But even if they were not, one should be aware that form 
factors, which are inevitably present, will serve to damp out bare integrals; 
see section VI.}. Hence there are no adjustable parameters in our scheme. Note 
that we do not consider gluon corrections and resultant `penguin diagrams' in 
the first place, since gluons are already assumed to be incorporated into 
dressing the bare quarks and producing the constituent fields which then 
interact via simple higher symmetry rules.

The layout of the paper is as follows. We summarise the working rules for
strong interactions and their weak perturbations in the next section. Then
we characterise the types of diagram that should be calculated in the
following section. It turns out that there are three types and they are 
evaluated one by one in Sections IV, V and VI. Next we see to what extent the
results are altered by introducing form factors into the weak interactions
(Section VII). Finally we apply the ideas to typical pseudoscalar mesons 
decays, to check whether we are on the right track; we find that the top quark 
contributions especially are overestimated unless the couplings are scaled
down by $1/\sqrt{M}$ factors where $M$ is the mass of the heavy quark, as 
indicated by heavy quark symmetry. With such damping included, many of the
amplitudes fall in the right ballpark---within a factor of two or three. The 
ability of the scheme to reproduce the `$\Delta I =1/2$ rule' in most cases is 
an encouraging sign. However having said that, there are a number of cases
in which the supermultiplet predictions are dreadful, indicating that we have 
overlooked some important feature or that our approach is fundamentally awry:
these are processes involving the decays of D-mesons or B-mesons to K-mesons. 
It is entirely possible that for such heavy/strange objects our hope of
absorbing all soft gluon effects into the dressing of the quarks is misplaced, 
and that they play a very significant role in modifying the weak vertices
themselves. If this puzzle can be unravelled, the prospects for calculating 
other nonleptonic amplitudes, such as initial heavy vector mesons or baryons, 
are good.

\section{Strong and Electroweak Interactions}  

The starting point of our model is that hadrons are reasonably well described
by multiquark, constituent composites with supermultiplet wavefunctions
$A=\alpha a$, where $\alpha$ is a Dirac spinor label and $a$ stands for
flavour:
\begin{equation}
 \Phi(p)_A^B = [(1+\gamma.v)(\gamma_5P +\gamma^\mu V_\mu)]_A^B;\quad
 m=m_a+m_b,\,p=mv
\end{equation}
\begin{eqnarray}
 \Psi_{(ABC)}(p)&=&[(1+\gamma.v)\gamma^\mu C]_{\alpha\beta}u_{\mu(abc)}(p)+
  \nonumber \\
   & &\{[(1+\gamma.v)\gamma_5C]_{\alpha\beta}\epsilon_{abd}u^d_{c\gamma}(p)
          +{\rm perms} \};\quad m=m_a+m_b+m_c,\,p=mv
\end{eqnarray}
of ground state mesons ($0^-, 1^-$) and baryons ($1/2^+, 3/2^+$), with 
tree-level interactions given by momentum-conserving effective Lagrangians like,
\begin{equation}
 g\bar{\Psi}^{(ABC)}\Psi_{(ABD)}\Phi_C^D,\quad f\Phi_A^B\Phi_B^C\Phi_C^A.
\end{equation}
Such algebraic contractions of indices can be represented pictorially as
a joining of the flavour quantum numbers via duality diagrams, with the
remaining contractions over spinorial indices serving to provide the
Lorentz structure of the amplitude, in keeping with higher symmetry
requirements. In (3), with the normalizations used in (1) and (2), the 
coupling constant $g$ is dimensionless while $f$ has dimensions of mass; we 
shall return to this point presently.

Now consider the addition of electroweak perturbations on the above sorts
of effective Lagrangians, by invoking the standard electroweak Lagrangian,
\begin{equation}
{\cal L}=\sum_{\rm flavour}\bar{\psi}[\gamma.(i\partial+{\cal W})-g_HH]\psi
       - \vec{W}^{\mu\nu}\cdot\vec{W}_{\mu\nu}/4 + {\cal L}_H + \ldots,
\end{equation}
where the supermultiplet boson field is
\begin{equation}
{\cal W} = -eQA + g_W VW(1-i\gamma_5)/\sqrt{8} + \ldots; \quad
             V = {\rm CKM~matrix},\,\,Q = {\rm charge}.
\end{equation}
By using the Particle Data Group parametrization ($c_{ij}\equiv\cos\theta_{ij},
s_{ij}\equiv \sin\theta_{ij}$),
\begin{equation}
\left( \begin{array}{ccc}
        V_{ud} & V_{us} & V_{ub}\\
        V_{cd} & V_{cs} & V_{cb}\\
        V_{td} & V_{ts} & V_{tb} 
       \end{array} \right)\! =\!
\left( \begin{array}{ccc}
        c_{12}c_{13} & s_{12}c_{13} & s_{13}\exp(-i\delta)\\
        -s_{12}c_{23}-c_{12}s_{23}s_{13}\exp(i\delta) & c_{12}c_{23}-
  s_{12}s_{23}s_{13}\exp(i\delta) & s_{23}c_{13}\\
  s_{12}s_{23}-c_{12}c_{23}s_{13}\exp(i\delta) & -c_{12}s_{23}-
  s_{12}c_{23}s_{13}\exp(i\delta) & c_{23}c_{13} 
       \end{array} \right),
\end{equation}
one guarantees that unitarity is respected and is not subject to errors by
picking specific values for certain matrix elements $V_{ij}$, without regard
to other elements. In practice we shall use the `average' values, $\theta_{12}=
0.223,\theta_{13}=0.0031,\theta_{23}=-0.039$ and disregard CP-violation effects
which lie outside the scope of this article, by setting $\delta = 0$.
In (4) we have not bothered 
to incorporate the Z-interactions because our main concern is flavour-changing 
processes. The most significant aspect of (4) is that it is normally applied 
to current quark fields, not the more massive constituent quark fields which 
are dynamically induced via strong interactions. It becomes questionable 
then whether (5) and (6) are applicable to the effective quark fields 
present in (3). Indeed previous experience shows that we must anticipate 
nontrivial corrections of around 30\% due to renormalization effects; 
for instance the chiral interaction should be changed from the left-projection 
$P_L=(1 - i\gamma_5)/2$ to about $(1-\frac{3}{4}\gamma_5)/2$ so that the 
axial vector component for the nucleon is reduced from 5/3 (arising the higher 
symmetry D/F ratio of 3/2) to the experimental value of $g_A/g_V\simeq 5/4$. 
The principal goal of this paper is to see if one can get sensible estimates 
of {\em all} nonleptonic amplitudes without invoking extra parameters, so we 
will ignore relatively small renormalization effects on the axial current 
but not the more substantial dependence on the mass of the participating
constituent quarks in the hadrons. Refinements can come later.

Flavour-changing decay amplitudes are governed by virtual W-exchange 
between the quarks. We can recognise three types of contribution, where
\begin{itemize}
 \item a charge-conserving transition occurs on one of the quark lines, 
 either at the start or at the finish (wave-function-like renormalizations) 
 or via a vertex-like correction; see Figures 1a,b,c.
 \item the charge exchange takes place between the participating quarks in 
 the hadron, again either as a self-energy or a pair of vertex corrections;
 see Figures 2a,b,c. (If these participating quarks comprise a meson, it 
 must be uncharged.)
 \item the W acts as a quark annihilation intermediary (see Figure 3)
 and is just what one encounters in semileptonic processes. It requires a
 charged meson to be one of the participating particles, of course.
\end{itemize}
For some processes, there might be a missing type; for example there is no 
annihilation diagram in K$^0\rightarrow \pi^0\pi^0$ decay.

Since there is a well-defined prescription for treating the strong  
vertex---where the quark tramlines join up---it only remains to estimate
the Feynman integrals corresponding to the W-loop exchange. This we shall
presently do, using the Feynman-'tHooft gauge field propagator,
$$\Delta_{\mu\nu}(k) = -\eta_{\mu\nu}/(k^2-m_W^2).$$
The left matrix $\gamma_{L\mu}=\gamma_\mu(1-i\gamma_5)/2$ multiplying
the propagator has a number of properties which come in handy during the 
calculations; aside from the obvious utility of the left-handed projection, 
there is the bonus that Fierz identities can be used to shuffle 
$\gamma_L\otimes\gamma_L$ from one pair of fermion lines into another pair, 
thereby relating disparate amplitudes.

This paper will concentrate on the pseudoscalar mesons decays, to verify
if the ideas have any sort of validity. We do not seek complete accuracy 
to any number of decimal places but will content ourselves if we can 
capture most if not all the amplitudes to within about 30\% or better. 
(If the scheme is successful, the generalisation to vector meson and baryon 
non-leptonic decays presents itself as the next obvious step.) Throughout we
will track flavour indices by making great use of quark line diagrams; sewing
the Dirac spinorial indices for each diagram then provides the Lorentz 
structure. Now, disregarding weak interactions, the three-meson vertex is 
fixed by the coupling factor $f$ and, because $0^-$ is forbidden to decay 
into two other pseudoscalars via strong interactions, one must look to the
strong decay $\rho\rightarrow 2\pi$, say, in order to get the value of $f$.

Two quark line diagrams determine $g_{\rho\pi\pi}$ and each of them contains 
a flavour normalization factor $1/\sqrt{2}$, but with opposite sign. 
Referring to Figure 4, with all momenta $p_i$ taken as incoming, one 
therefore arrives at the effective interaction Lagrangian,
\begin{eqnarray*}
 {\cal L}_{\rho\pi\pi}&=&f\,{\rm Tr}[\Phi_P(p_3)\Phi_P(p_2)\Phi_V(p_1)
           /\sqrt{2} -  (1 \leftrightarrow 2)]/4 \nonumber \\
 &=&f\,{\rm Tr}[(1+\gamma.v_3)\gamma_5(1+\gamma.v_2)\gamma_5(1+\gamma.v_1)
        \gamma.\epsilon_1-(1\leftrightarrow 2)]/4\sqrt{2} \nonumber \\
 &=&(p_3-p_2).\epsilon_1 f(m_1+m_2+m_3)/\sqrt{2}m_2m_3 
    \equiv g_{\rho\pi\pi}(p_3-p_2).\epsilon_1
\end{eqnarray*}
Experimentally, the $\rho$-decay width tells us that the dimensionless
coupling $g_{\rho\pi\pi} \simeq 6.03$. Thus $f(6\hat{m})/4\hat{m}^2 = 
\sqrt{2}g_{\rho\pi\pi}$, or $f=\sqrt{8}\hat{m}g_{\rho\pi\pi}/3 \sim 2$ GeV.
Now $f(m_1+m_2+m_3)$ is a ubiquitous factor in relativistic supermultiplet
theory, so we shall adopt this quantity as the `universal' value which 
equals about 4.1 GeV$^2$, based on the assumption that the quark masses are 
not far from $m_u\simeq m_d \equiv \hat{m} \sim 0.34$ GeV. 

\section{Type I - Quark Line Transitions} 

First we shall deal with changes of flavour (but not of charge) on a single 
quark, which can take place as self-energy-like diagrams at each of the
hadronic legs (Figures 1a, 1c) or as a vertex correction across the legs
(Figure 1b). In both cases, one must sum over the internal quark flavours;
a generic case (Delbourgo and Scadron 1985; Fuchs and Scadron 1986) is the 
$s-d$ transition, where we have to sum over $u,c,t$. Quite generally, between 
$i$ and $k$ quarks, the self-energy part is 
$\Sigma_{ik}(p) \equiv \sum_jV_i^jV_k^{j*}\Sigma_j(p)$ where
\begin{equation}
 \Sigma_j(p) =
 i\frac{g_W^2}{2} \int \frac{\bar{d}^4k}{k^2-m_W^2}\gamma_{L\mu}
  \frac{\eta^{\mu\nu}}{\gamma.(p+k)-m_j} \gamma_{L\nu}
 \equiv p.\gamma_L {\cal F}_j(p^2,m_W,m_j).
\end{equation}
Now ${\cal F}$ is potentially troublesome because, {\em neglecting form
factors}, it contains a logarithmic divergence:
\begin{equation}
 {\cal F}_j = -i\frac{g_W^2}{p^2}\int \frac{\bar{d}^4k\,\,p.(p+k)}
              {[k^2-m_W^2][(k+p)^2-m_j^2]}
        =\int_0^1\!\!dx\int \frac{(1-x)\,\bar{d}^4k}{[k^2+p^2x(1-x)-m_j^2x
                                     -m_W^2(1-x)]^2}.
\end{equation}
However, through unitarity of the CKM matrix and the fact that we are
studying flavour changing transitions ($i\neq k$), we can subtract off the
dangerous divergent part, contained in the limit $m_j\rightarrow 0$, and 
happily use the convergent difference instead of ${\cal F}$:
$${\cal F}_j(p^2,m_W,m_j)-{\cal F}_j(p^2,m_W,0)=-\frac{g_W^2}{16\pi^2}
 \int_0^1dx\,(1-x)\ln\left[1 + \frac{m_j^2 x}{(1-x)(m_W^2-p^2x)}\right].$$
This can then be suitably approximated; thus for $p^2 \ll m_W^2$ which 
applies to all {\em external} hadrons not containing a top quark,
\begin{equation}
 {\cal F}_j(m_j)-{\cal F}_j(0)\!\simeq\!
  -\frac{g_W^2}{16\pi^2}\int_0^1\!\! dx\,\,x\ln\left[1+\frac{m_j^2(1-x)}
    {xm_W^2}\right]\!=\! -\frac{g_W^2}{32\pi^2}\frac{J}{1-J}
   \left[1 + \frac{J\ln J}{1-J}\right];\,J\equiv \frac{m_j^2}{m_W^2}.
\end{equation}
In particular, for $m_j\ll m_W$ or $J\ll 1$, we get a very good estimate
of (9) with 
\begin{equation}
 {\cal F}_j(m_j)-{\cal F}_j(0) \simeq
-\frac{g_W^2J}{32\pi^2}=-\frac{1}{2}\left(\frac{g_Wm_j}{4\pi m_W}\right)^2.
\end{equation}
But for the top quark one must be more careful in approximating (9). Taking
the experimental value of $m_t$ and $m_W$ as inputs we estimate
$${\cal F}_t(m_t)\!-\!{\cal F}_t(0)\simeq -\frac{1.23g_W^2}{32\pi^2}\simeq
 -\frac{0.13g_W^2m_t^2}{16\pi^2m_W^2},$$
which is almost exactly 1/4 of the expression (10). Remembering that 
$G_F/\sqrt{2} = g_W^2/8m_W^2$ = 8.25$\times 10^{-6}$ GeV$^{-2}$, we may 
finally write the transition element,
\begin{equation}
\Sigma_{ik}(p) \equiv p.\gamma_L{\cal F}_{ik}
 = -\frac{G_Fp.\gamma_L}{4\sqrt{2}\pi^2}\sum_j V_i^jV_k^{j*}m_j^2\rho_j,
\end{equation}
where the weight factor $\rho_q$=1 for all but the top quark, when $\rho_t
\simeq 1/4$.

Evaluation of the transition factors, ${\cal F}_{ik}=-\sum_j V_i^jV_k^{j*}
\,G_Fm_j^2\rho_j/4\sqrt{2}\pi^2$, is reasonably straightforward, using
fairly well-known values of mixing angles and the (GeV) values $m_u\simeq m_d 
\simeq 0.34$,$m_s\simeq 0.48$, $m_c\simeq 1.5$, $m_b\simeq 4.7$, $m_t\simeq 
175$. The values are listed in the left-hand columns of Table I, and it should 
be noted that transition elements for down-type quarks depend significantly on 
the contribution from the intermediate top quark; the effect is smallest for 
the $s-d$ transition, but even so the $t$ quark competes well with the 
$u,c$ contributions; mostly it dominates the other contributions, in spite 
of the fact that off-diagonal $V_t^q$ terms are quite small.

Next we turn to Figure 1b, which produces the (matrix) vertex integral,
\begin{eqnarray}
\Gamma_{ik}&=&-i\sum_jV_i^jV_k^{j*}\frac{g_W^2}{2}\int
    \frac{\bar{d}^4k}{k^2-m_W^2}\gamma_{\mu L}
    \frac{1}{\gamma.(p_k+k)-m_j}\frac{1}{\gamma.(p_i+k)-m_j}
    \gamma^\mu_L \nonumber \\
&=&ig_W^2\sum_jV_i^jV_k^{j*}\int\frac{m_j\,(p_k+p_i+2k).\gamma_L\,\bar{d}^4k}
     {[(p_k+k)^2-m_j^2][k^2-m_W^2][(p_i+k)^2-m_j^2]}.
\end{eqnarray}
where $p_i$ is the incoming momentum of the $i$-quark and $p_k$ is the
outgoing momentum of the $k$-quark. Introducing Feynman parameters and 
assuming that $p_{i,k}\ll m_W$ ---as is true for all top-free external 
hadrons---we may get the one-dimensional parametric representation,
\begin{eqnarray}
\Gamma_{ik}&=&\sum_jV_i^jV_k^{j*}m_j(p_i+p_k).\gamma_L\frac{g_W^2}{16\pi^2}
  \int_0^1 \frac{x(1-x)\,\,dx}{m_W^2x+m_j^2(1-x)} \nonumber \\
      &=&\sum_jV_i^jV_k^{j*}m_j(p_i+p_k).\gamma_L\frac{g_W^2}{16\pi^2m_W^2}
       \left[\frac{J\ln{J}}{(1-J)^3}+\frac{1+J}{2(1-J)^2}\right];\quad
      J\equiv\frac{m_j^2}{m_W^2}.
\end{eqnarray}
When the intermediate $j$-quark is considerably lighter than the W, we can 
make the reasonable approximation, $J\ll 1$ and find
$$\Gamma_{ik} \simeq \frac{1}{2}\left(\frac{g_W}{4\pi m_W}\right)^2
            (p_i+p_k).\gamma_L\,\sum_j V_i^jV_k^{j*}m_j, $$
but for the top quark, the numerical value of $J \sim 4.7$, means that
the contribution is about 1/8 of what the light quark approximation above
gives. We shall therefore write the vertex correction in the form
\begin{equation}
 \Gamma_{ik}\equiv \frac{1}{2}(p_i+p_k).\gamma_L{\cal G}_{ik} =
 \frac{G_F(p_i+p_k).\gamma_L}{4\sqrt{2}\pi^2}\sum_jV_i^jV_k^{j*}m_j\sigma_j,
\end{equation}
where the weight factor $\sigma_q =1$ for all quarks but the top, when 
$\sigma_t\simeq 1/8$. The magnitudes of the vertex transition elements 
${\cal G}_{ik}$ are listed in the second columns of Table I; the effect from 
the top is not so significant as for the self-energy elements, but it still
dominates the $b\leftrightarrow s,d$ transitions.

We should point out that the above results were derived on the assumption
that the W-coupling to the constituent quarks has no form factors; but
this assumption is obviously not correct. A more refined calculation ought to 
include form factors of the pole type $F(k^2)\sim M^2/(M^2-k^2)$ or something 
fancier. We shall return to this point later.

It remains to sum up the terms arising from Figures 1a, 1b and 1c. Since
hadronic supermultiplets consist of constituent quarks on their mass shells
sharing the total momentum according to their mass (they all have equal
velocity) with negligible binding, we shall evaluate the result between 
free spinors $\bar{u}(p_k)..u(p_i)$. In doing so we must be careful to
{\em halve} the self-energy contributions on external quark lines, since
they are eventually associated with $Z^{1/2}$ renormalization factors.
Hence, between spinors, the sum equals
\begin{eqnarray*}
{\cal T}^I_{ik}&=&-\frac{1}{2(\gamma.p_i-m_k)}\Sigma_{ik}(p_i)+
 \Gamma_{ik}(p_i,p_k)-\Sigma_{ik}(p_k)\frac{1}{2(\gamma.p_k-m_i)}\nonumber\\
&=&-\frac{{\cal F}_{ik}}{2(\gamma.p_i-m_k)}p_i.\gamma_L+
   \frac{1}{2}(p_i+p_k).\gamma_L{\cal G}_{ik}
   -p_k.\gamma_L \frac{{\cal F}_{ik}}{2(\gamma.p_k-m_i)} \nonumber \\
&=&-\frac{{\cal F}_{ik}}{4}\left[1 + i\gamma_5\frac{m_k-m_i}{m_k+m_i}\right]
   + \frac{{\cal G}_{ik}}{4}\left[(m_k+m_i) + i\gamma_5(m_i-m_k)\right].
\end{eqnarray*}

The last step is to contract ${\cal T}^I$ over the hadronic wavefunctions. 
For three 0$^-$ mesons, the only relevant part of ${\cal T}^I$ is the one 
containing $\gamma_5$ so as to get a non-vanishing trace. Therefore,
with the flavour labels of Figure 1, we get the generic amplitude:
\begin{eqnarray}
 M^I_{sdcu}&=&f\,{\rm Tr}[\Phi_P(p_2){\cal T}^I\Phi_P(p_1)\Phi_P(p_3)]
 /m_1m_2m_3 \nonumber \\
 &=&i\frac{(m_d-m_s)f}{4m_1m_2m_3}\left[{\cal G}_{sd}+
  \frac{{\cal F}_{sd}}{m_s+m_d}\right]
  {\rm Tr}[(\gamma.p_2+m_2)(\gamma.p_1+m_1)(\gamma.p_3-m_3)]\nonumber\\
 &=&i\frac{f(m_1+m_2+m_3)(m_s-m_d)[m_3^2-(m_1-m_2)^2]}{2m_1m_2m_3}
 \left[{\cal G}_{sd} + \frac{{\cal F}_{sd}}{m_s+m_d}\right].
\end{eqnarray}
Recall that the $m_i$ in (14) are the sums of the constituent quark masses
comprising the hadron. When analysing any other flavour changing amplitude 
of type I, it is a simple matter of substituting the appropriate flavour 
labels in (14) above. We shall frequently be doing so in section VI.

\section{Type II - W-Exchange across quarks}

We now turn to the graphs contained in Figure 2. The first of these 
actually corresponds to a $u\bar{c}$ transition into two mesons, dominated
by an intermediate $d\bar{s}$ state. Neglecting the small momentum
carried by the W meson relative to its mass, this particular contribution
is given by the generic contraction,
\begin{eqnarray}
 M^{IIA}_{udbsc}&=&-\frac{g_W^2\,f}{32m_W^2}V_u^dV_c^{s*}{\rm Tr}
 [(\Phi_P(p_2)(\gamma.p_d\!+\!m_d)
 \gamma_{L\mu}\Phi_P(p_1)\gamma_L^\mu(\gamma.p_s\!+\!m_s)\Phi_P(p_3)]\nonumber \\
 &=&\frac{g_W^2m_dm_sfV_d^uV_s^{c*}}{32m_W^2m_1m_2m_3}{\rm Tr}
 \left[(\gamma.p_2\!-\!m_2)\left(1\!+\!\frac{\gamma.p_1}{m_s\!+\!m_d}\right)
 i\gamma_L.p_1\left(1\!-\!\frac{\gamma.p_1}{m_s\!+\!m_d}\right)
 (\gamma.p_3\!+\!m_3)  \right]
 \nonumber \\
 &=&\frac{iG_Ff(m_1\!+\!m_2\!+\!m_3)m_dm_sV_u^dV_c^{s*}}
 {2\sqrt{2}m_1m_2m_3}\left[1\!-\!\left(\frac{m_u\!\!+\!m_c}{m_s\!\!+\!m_d}
 \right)^2\right](m_d\!\!-\!m_s)(m_1\!-\!m_2\!-\!m_3),
\end{eqnarray}
because in the intermediate state, $p_d=p_1m_d/M_1, p_s=-p_1m_s/M_1; 
M_1=m_s+m_d$.

Competing with this answer are the vertex corrections of Figures 2b,2c.
The latter involve Feynman integrals which are more difficult to calculate
analytically. An interesting technical aspect of the evaluation is that
they contain cancelling logarithmic divergences, irrespective of CKM
matrix unitarity. Figure 2b and 2c yield, respectively
\begin{equation}
M^{IIB}_{udbsc}\!=\!-i\frac{g_W^2f}{8}\int\frac{V_u^dV_c^{s*}\,\bar{d}^4k}
 {k^2-m_W^2}{\rm Tr}\left[\Phi_P(p_2)\frac{1}
 {\gamma.(p_u\!+\!k)\!-\!m_d}\gamma_{L\mu}\Phi_P(p_1)
 \frac{1}{\gamma.(p_s\!-\!k)\!-\!m_c}\gamma_L^\mu\Phi_P(p_3)\right],
\end{equation}
\begin{equation}
M^{IIC}_{udbsc}\!=\!-i\frac{g_W^2f}{8}\int\frac{V_u^dV_c^{s*}\,\bar{d}^4k}
 {k^2-m_W^2}{\rm Tr}\left[\Phi_P(p_2)\gamma_{L\mu}
 \frac{1}{\gamma.(p_d\!-\!k)\!-\!m_u}\Phi_P(p_1)\gamma_L^\mu
 \frac{1}{\gamma.(p_c\!+\!k)\!-\!m_s}\Phi_P(p_3)\right],
\end{equation}
where now $p_s=m_sp_3/m_3, p_u=m_up_1/m_1, p_d=-m_dp_2/m_2, p_c=-m_cp_1/m_1$.
Simplifying the sum,
\begin{eqnarray*}
M^{IIBC}_{udbsc}&=&\frac{g_W^2f}{8}\int\frac{V_u^dV_c^{s*}\,\bar{d}^4k}
    {k^2-m_W^2}\,\,{\rm Tr}\nonumber \\
   & &\left[\Phi_P(p_2)\!\!\left(\frac{1}{\gamma.(p_u\!+\!k)\!-\!m_d}
     \frac{\gamma.(k\!-\!p_c)\!-\!m_s}{(k\!-\!p_s)^2\!-\!m_c^2} -
     \frac{\gamma.(k\!-\!p_u)\!-\!m_d}{(k\!-\!p_d)^2\!-\!m_u^2}
     \frac{1}{\gamma.(k\!+\!p_c)\!-\!m_s}\right)\!\!\Phi_P(p_3) \right].
\end{eqnarray*}

This calculation is messy---suggesting numerical methods as a last 
resort---if all external momenta are religiously kept within the 
Feynman integral. However, we can achieve an reasonable estimate of
the result by going to the soft limit, i.e. neglecting the $p$-dependence
within the propagators, relative to the large momentum $k$ carried by
the W line. Using the supplementary integral,
\begin{eqnarray*}
\int\frac{i\,\bar{d}^4k}{k^2\!-\!m_W^2}\left[
 \frac{1}{(k\!-\!p_u)^2\!-\!m_d^2}\!\right.&-&
 \left.\!\frac{1}{(k\!-\!p_c)^2\!-\!m_s^2}\right]
 =\int_0^1\frac{dx}{16\pi^2}\ln\left[
 \frac{m_d^2x\!+\!m_W^2(1\!-\!x)\!-\!m_u^2x(1\!-\!x)}
   {m_s^2x\!+\!m_W^2(1\!-\!x)\!-\!m_c^2x(1\!-\!x)}\right]
  \nonumber \\
 &\simeq&\frac{1}{32\pi^2m_W^2}\left[m_c^2+2m_s^2\ln\left(\frac{m_s^2}{m_W^2}
       \right)-m_u^2-2m_d^2\ln\left(\frac{m_d^2}{m_W^2}\right)\right],
\end{eqnarray*}
valid when $m_i^2\ll m_W^2$, and keeping leading logarithms,
 one may estimate
\begin{eqnarray*}
M^{IIBC}_{udbsc}\simeq\frac{g_W^2f}{8}\!\int\!\bar{d}^4k& &V_u^dV_c^{s*}
    \frac{k^2-m_dm_s}{k^2-m_W^2}\\
 &&{\rm Tr}\left[\Phi_P(p_2)\left(\frac{1}{(k^2-m_c^2)(k^2-m_d^2)}-
  \frac{1}{(k^2-m_u^2)(k^2-m_s^2)}\right)\Phi_P(p_3)\right]
\end{eqnarray*}
\begin{eqnarray}
=-\frac{iG_FV_u^dV_c^{s*}\,f}{16\sqrt{2}\pi^2m_2m_3}
  &&\left[m_1^2-(m_2+m_3)^2\right]  \nonumber\\
  &&\left[(m_u^2-m_c^2)\ln\left(\frac{m_c^2m_u^2}{m_W^4}\right)
+\frac{1}{2}(m_s^2-m_d^2)\ln\left(
 \frac{m_s^2m_d^2m_c^2m_u^2}{m_W^8} \right) \right].
\end{eqnarray}
The significant point about this last result is that it is of the same order
of magnitude as $M^{IIA}$; even though there is a suppression factor of
$1/4\pi^2$ from the integration in eq (18), it is compensated by a number of
logarithms which are typically in the range $\ln(m_W^2/m_s^2) \sim 10$. In 
fact, from our viewpoint, cancellations between these sorts of terms are 
responsible for the small size of the $\Delta I = 3/2$ amplitude in K decays.

\section{TYPE III - W-Annihilation}
Evidently, this process only applies to charged mesons, which may be incoming
or outgoing. A typical example is drawn in Figure 3. It is easy to calculate,
from what has gone before. Basically, we take advantage of the fact that 
through Fierz reshuffling,
$(\bar{u}_1\gamma_L^\mu u_2).(\bar{u}_3\gamma_{L\mu} u_4)=
-(\bar{u}_3\gamma_L^\mu u_2).(\bar{u}_1\gamma_{L\mu} u_4)$, aside 
from colour factors; and because we have an extra fermion loop in Figure 3 
relative to Figure 2a, the extra (-) sign is effectively swallowed up. 
Thus, on the assumption that constituent quarks interact left-handedly to a 
first approximation, Figure 3 gives the same answer as Figure 2a so far as 
the Lorentz contraction is concerned. Transcribing the flavour labels, 
and incorporating a colour factor of 3, we get the generic amplitude,
\begin{equation}
M^{III}_{sdbcu}=\frac{3iG_Ff(m_1\!+\!m_2\!+\!m_3)m_dm_cV_u^sV_c^{d*}}
 {2\sqrt{2}m_1m_2m_3}\left[1\!-\!\left(\frac{m_s\!\!+\!m_u}{m_c\!\!+\!m_d}
 \right)^2\right](m_d\!\!-\!m_c)(m_1\!-\!m_2\!-\!m_3).
\end{equation}

\section{Form factor corrections}

The integrals in sections III-V were derived on the assumption that the
coupling of the W-boson to the quarks was pointlike. In fact the weak current
must be affected by intermediate vector and pseudoscalar bosons that can
latch on to the quark fields and this will produce a natural damping. 
(This phenomenon is very familiar in QED and is responsible for the finiteness
of photonic corrections to the proton neutron mass difference, aside from the 
contribution due to the intrinsic $u-d$ mass difference. It is typically 
governed by a strong interaction scale of about 1.1 GeV.) We may estimate the 
effects of mediating mesons by incorporating the form factor $M^2/(M^2-k^2)$
at each W-leg, where $M$ is some kind of geometric mean of the intermediate 
mesons on each side of the W-line. The consequence is that the integral (8) say
is finite, regardless of CKM unitarity, because it gets modified to
\begin{eqnarray}
{\cal F}_j &=& -i\frac{g_W^2}{p^2}\int \frac{\bar{d}^4k\,\,p.(p+k)}
              {[k^2-m_W^2][(k+p)^2-m_j^2]}\left(\frac{M^2}{M^2-k^2}\right)^2
         \nonumber \\
&=& \frac{g_W^2L^2}{32\pi^2}\left[\frac{1}{1-L}\left(
    \frac{1}{L}+\frac{\log L}{1-L}+\frac{J^2\log J}{(L-1)(J-1)^2}\right)\right.
         +\nonumber \\
& & \qquad\quad\left.\frac{J}{(L-J)^2}\left(\frac{1}{J-1}+\frac{J}{L(L-1)}+
     \frac{J\log(J/L)}{(1-L)^2}+\frac{2J\log J}{(L-1)(L-Q)}\right)\right],
\end{eqnarray}
where $L\equiv M^2/m_W^2, J\equiv m_J^2/m_W^2$. The main effect is to enhance
the contribution from the top quark; this can be quite substantial and having
included it, we must rescale the universal constant $f$ down appropriately so 
that the plain results (no form factors), which should be governed by $G_F$, do
not go badly out of line. Once this is done there is no more room for maneouvre.

The same sort of modification arises in the vertex integral ${\cal G}$ but 
we shall not bother to exhibit the dependence on $L$ and $J$ because the 
consequences are very minor; the point is that each ${\cal G}$ contribution
is given by a well-behaved finite integral. The form factor effects here, in
contrast to those on ${\cal F}_t$, are very tame and amount to corrections of 
just a few per cent.

\section{Typical Results and Difficulties}

Let us now describe some of the consequences of the supermultiplet scheme and
the ensuing problems. To keep the discussion clean, we shall ignore channels
which involve uncharged mesons that can mix, like $\eta$ and $\eta'$, so we
will focus on decays where the outgoing particles are pions, kaons and heavy
mesons like D and D$_s$. Below we define $M$ to be the magnitude of the
decay amplitude (it has dimensions of mass), which is derived from the partial 
decay width via
$$\Gamma_{m\rightarrow m_1m_2} = |M|^2\Delta/16\pi m^3;\qquad \Delta \equiv
                                \sqrt{[m^2-(m_1-m_2)^2][m^2-(m_1+m_2)^2]}.$$
Throughout we have disregarded the $u-d$ mass difference and used Mathematica
to compute the integrals numerically, as required.

First let us consider the time-honoured example of kaon decay. Two typical 
cases are K$^+ \rightarrow \pi^+\pi^0$, corresponding to $\Delta I =3/2$, and
K$_s \rightarrow \pi^+\pi^-$ which amounts to $\Delta I =1/2$. It is of course
known that $M^{expt}_{K_s\pi^+\pi^-} = 3.91\times 10^{-7}$ is about twenty
times larger than $M^{expt}_{K^+\pi^+\pi^-} = 1.83\times 10^{-8}$ and this
is the main feature to be `explained'. When we tackle these cases via the 
supermultiplet scheme, we find that with the longer-lived K$^+$, the type I
contributions cancel, as they must, but for the short-lived K$^0$ there
is a significant type I contribution (because the $s-d$ transition is
$\Delta I=1/2$); thus
$$M_{K_s^0\rightarrow \pi^+\pi^0}=\sqrt{2}[M^I_{sdud}+M^{II}_{sudud}+
      M^{III}_{uudsd}]$$
$$M_{K^+\rightarrow\pi^+\pi^0}=[M^{II}_{usudu}+M^{III}_{usuud}]/\sqrt{2}$$
have a ratio of about 4, which is still a factor of 5 too small; but this
is easily remedied by including form factors\footnote{and renormalizing the 
strong coupling because of the enhancement factors.} using a weak cutoff of 
about 1.25$m_t$. In this way we can obtain values which are rather close to 
experiment:
$$|M_{K_s^0\rightarrow \pi^+\pi^0}|\simeq 3.9\times 10^{-7}{\rm~GeV};\quad
  |M_{K^+\rightarrow\pi^+\pi^0}|\simeq 1.9\times 10^{-8}{\rm~GeV}.$$

If one turns to D-meson decays, there are some good results, but there are
also some dreadful ones. For example, we get the nice answer
$$|M_{D^0\rightarrow \bar{K}^0\pi^0}|=|M^{II}_{duucs}-M^{II}_{csddu}|/\sqrt{2}
 \simeq 1.79\times 10^{-6} {\rm~GeV};\quad {\rm cf~} 
  |M_{D^0\rightarrow\bar{K}^0\pi^0}^{\rm expt}|\simeq 
  1.85\times 10^{-6} {\rm~GeV},$$
as well as the ridiculous value (in GeV)
$$|M_{D^+\rightarrow \pi^0\pi^+}|=|M^{I}_{cuud}-M^{II}_{dudcd}-M^{III}_{dsdcu}|
 /\sqrt{2} \simeq 6.08\times 10^{-5}; \,{\rm cf~}
   |M_{D^+\rightarrow \bar{K}^0\pi^+}^{\rm expt}|\simeq 
   1.35\times 10^{-6}.$$
In fact most of the results involving heavy mesons are predicted to be too
large in the raw supermultiplet scheme; it is not hard to trace the reason for 
this effect and thereby cure it.

The point is that the supermultiplet interactions which we wrote down previously
did not take account of the $1\sqrt{M}$ diminution of matrix elements which are
expected from heavy quark theory, in order to give symmetry results at equal 
velocity, {\em not equal momentum}. These factors have little effect on the
light quark composites but play a substantial role for the heavy mesons. When 
the appropriate factors $\sqrt{2m_u/(m_i+m_k)}$ for a meson composed of
quarks $i$ and $k$ are incorporated, they depress the nonleptonic decay
amplitudes of heavy mesons; thus $M_{D^+\rightarrow\pi^0\pi^+}$ goes down to
the acceptable value 1.14$\times 10^{-6}$ GeV and likewise for many other
matrix elements. Nevertheless there remain some processes whose reduction is
excessive, such as
$$|M_{D_s^+\rightarrow K^0\bar{K}^0}|=|M^{II}_{duscs}+M^{III}_{cusds}|\simeq  
  4.3\times 10^{-9}{\rm~GeV};\,{\rm cf~}
  |M_{D_s^+\rightarrow K^0\bar{K}^0}^{\rm expt}|=2.41\times 10^{-6}{\rm~GeV},$$
and others which stubbornly resist reduction, such as
$$|M_{B^+\rightarrow K^0\pi^+}|=|M^{I}_{sbdu}+M^{III}_{sbduu}|\simeq
 2.3\times 10^{-4}{\rm ~GeV};\,{\rm cf~}
 |M_{B^+\rightarrow K^0\pi^+}^{\rm expt}| \simeq 0.5\times 10^{-7}{\rm~GeV}.$$
The cure is therefore only partial and this is most disappointing.

In spite of a number of silly predictions, we believe that our approach makes 
sound philosophical sense even if the way we have applied the idea has not
been wholly successful: weak interactions should be regarded as a perturbation 
of the strong interactions and not the other way round. Maybe others will find 
errors with our numerical work and/or will be able to tackle the problem better
than we have. We have perhaps been over-ambitious in thinking that we could 
get away without introducing any parameters and it is possible that with the
introduction of many more couplings one can get suitable agreement with 
all the experimental results: the weak vertices do receive substantial
renormalization corrections from the strong interactions. To conclude on a
more optimistic note, if one could resolve the difficulties, it would be a
simple step to generalize our work to the vector mesons and the baryonic
supermultiplet.
\vspace{.3in}

\noindent {\bf Acknowledgments}
This work was supported by the Australian Research Council under grant
number  A69800907.
\vspace{.3in}

\noindent {\bf References}

\small
\noindent $^*$ Email: Bob.Delbourgo@utas.edu.au

\noindent $^{**}$ Email: D.Liu@utas.edu.au

\noindent Ali, A., Kramer, G., and Lu, C-D. (1998).{\em Phys. Rev.} {\bf D58},
 094009.
 
\noindent Ali, A., Kramer, G., and Lu, C-D. (1999).{\em Phys. Rev.} {\bf D59},
 014005.

\noindent Bauer, M., Stech, B., and Wirbel, M. (1987). {\em Z. Phys.} 
 {\bf C34}, 103.

\noindent Beneke, M. (2000) ``Conceptual aspects of QCD factorization in 
 hadronic B decays'', hep-ph/0009328.

\noindent Delbourgo, R., and Liu, D. (1996). {\em Phys. Rev.} {\bf D53}, 
 6576.

\noindent Delbourgo, R. and Liu, D. (1998). `` Nonleptonic decays: amplitude
 analysis and supermultiplet schemes'', pp 249-260 in `Nonperturbative Methods 
 in Quantum Field Theory', (ed. by A.W. Schreiber, A.G. Williams and A.W. 
 Thomas), (World Scientific, Singapore).

\noindent Delbourgo, R., and Scadron, M.D. (1985). {\em Nuovo Cim. Lett.} 
{\bf 44}, 193.

\noindent Fuchs, N.H., and Scadron, M.D. (1986). {\em Nuovo Cim.} {\bf A93}, 
 205.

\noindent Isgur, N., and Wise, M.B. (1989). {\em Phys. Lett.} {\bf B232}, 113.

\noindent Isgur, N., and Wise, M.B. (1990). {\em Phys. Lett.} {\bf B237}, 527.

\noindent Lu, C-D. (1999). {\em Nucl. Phys. Proc. Suppl.} {\bf 74}, 227.

\noindent Neubert, M. (1994). {\em Phys. Rept.} {\bf 245}, 259.

\noindent Neubert, M. (1997). ``B-decays and heavy quark expansion'' in `Heavy 
 Flavours II', pp 239-293 (World Scientific, Singapore).

\noindent Neubert, M. (2000). ``Application of QCD Factorization in Hadronic
B-Decays'', Cornell preprint.

\noindent Neubert, M., and Stech, B. (1997). ``Nonleptonic weak decays of 
 B-mesons'', in `Heavy Flavours II', pp 294-344 (World Scientific, Singapore).

\noindent Sakita, B., and Wali, K.C. (1965). {\em Phys. Rev.} {\bf 139}, B1355.

\noindent Salam, A., Delbourgo, R., and Strathdee, J. (1965). {\em Proc. R. 
 Soc. London}, {\bf A284}, 146.

\noindent Scadron, M.D. (1997). {\em Nuovo Cim.} {\bf 110A}, 865.

\noindent Stech, B. (1997). ``Twenty Beautiful Years of Bottom Physics'', 
 (Chicago Univ. Press), hep-ph/9709280.


\normalsize
\narrowtext
\begin{table}
\caption{Flavour changing self-energy transition elements estimated on the 
 basis of eq (10), all in GeV units. The combination ${\cal H}_{ij}
\equiv{\cal G}_{ij}+{\cal F}_{ij}/(m_i+m_j)$ is needed in eq (15).}
\begin{tabular}{||r|c||r|c||r|c||}
${\cal F}_{sd}$ & $7.91\times 10^{-7}$ &
 ${\cal G}_{sd}$ & $-1.08\times 10^{-7}$ &
  ${\cal H}_{sd}$ & $2.11\times 10^{-7}$ \\
${\cal F}_{sb}$ & $-5.96\times 10^{-5}$ &
 ${\cal G}_{sb}$ & $3.17\times 10^{-7}$ &
  ${\cal H}_{sb}$ & $-2.80\times 10^{-6}$ \\
${\cal F}_{db}$ & $1.86\times 10^{-5}$ &
 ${\cal G}_{db}$ & $-1.00\times 10^{-7}$ &
  ${\cal H}_{db}$ & $9.96\times 10^{-7}$ \\
${\cal F}_{uc}$ & $-4.30\times 10^{-9}$ &
 ${\cal G}_{uc}$ & $1.15\times 10^{-8}$ &
  ${\cal H}_{uc}$ & $2.29\times 10^{-9}$ \\
${\cal F}_{ut}$ & $-1.44\times 10^{-8}$ &
 ${\cal G}_{ut}$ & $6.08\times 10^{-9}$ &
  ${\cal H}_{ut}$ & $1.50\times 10^{-9}$ \\
${\cal F}_{ct}$ & $1.78\times 10^{-7}$ &
 ${\cal G}_{ct}$ & $-6.88\times 10^{-8}$ &
  ${\cal H}_{ct}$ & $-1.70\times 10^{-8}$
\end{tabular}
\label{table1}
\end{table}

\begin{figure}[b]
\begin{center}
\begin{picture}(460,100)(10,60)
\ArrowLine(12,146)(38,146) \Text(18,156)[]{$s$}
\Text(14,143)[]{\scriptsize{1}}
\Vertex(38,146){2.0}
\Line(38,146)(68,146)
\ArrowLine(92,138)(12,138)\Text(18,130)[]{$u$}
\PhotonArc(53,146)(15,0,180){2}{10}\Text(53,170)[]{$W$}
\ArrowLine(68,146)(130,146) \Text(125,156)[]{$d$}
\Text(128,143)[]{\scriptsize{2}}
\Vertex(68,146){2.0}
\ArrowLine(92,90)(92,138)\Text(84,100)[]{$u$}
\ArrowLine(130,138)(100,138)\Text(125,130)[]{$c$}
\ArrowLine(100,138)(100,90)\Text(110,100)[]{$c$}
\Text(96,94)[]{\scriptsize{3}}
 \Text(74,65)[]{\bf (a)}
\Text(143,142)[]{$+$}
\ArrowLine(153,146)(199,146) \Text(163,156)[]{$s$}
\Text(155,143)[]{\scriptsize{1}}
\Vertex(199,146){2.0}
\Line(199,146)(229,146)
\Vertex(229,146){2.0}
\ArrowLine(229,146)(265,146) \Text(257,156)[]{$d$}
\Text(263,143)[]{\scriptsize{2}}
\PhotonArc(214,146)(15,10,170){2}{10}\Text(214,170)[]{$W$}
\ArrowLine(210,138)(153,138)\Text(163,130)[]{$u$}
\ArrowLine(265,138)(218,138)\Text(257,130)[]{$c$}
\ArrowLine(210,90)(210,138)\Text(202,100)[]{$u$}
\ArrowLine(218,138)(218,90)\Text(226,100)[]{$c$}
\Text(215,94)[]{\scriptsize{3}}
\Text(214,65)[]{\bf (b)}
\Text(279,142)[]{$+$}
\ArrowLine(288,146)(360,146) \Text(298,156)[]{$s$}
\Text(290,143)[]{\scriptsize{1}}
\Vertex(360,146){2.0}
\Line(360,146)(390,146)
\Vertex(390,146){2.0}
\ArrowLine(390,146)(420,146) \Text(416,156)[]{$d$}
\Text(418,143)[]{\scriptsize{2}}
\PhotonArc(375,146)(15,10,170){2}{10}\Text(375,170)[]{$W$}
\ArrowLine(328,138)(288,138)\Text(298,130)[]{$u$}
\ArrowLine(328,90)(328,138)\Text(320,100)[]{$u$}
\ArrowLine(336,138)(336,90)\Text(346,100)[]{$c$}
\ArrowLine(420,138)(336,138)\Text(416,130)[]{$c$}
\Text(333,94)[]{\scriptsize{3}}
\Text(364,65)[]{\bf (c)}
\end{picture}
\end{center}
\caption{Single quark line transition.}
\end{figure}

\begin{figure}
\begin{center}
\begin{picture}(600,265)(10,-100)
\ArrowLine(12,146)(38,146) \Text(18,156)[]{$u$}
\Text(14,138)[]{\scriptsize{1}}
\Vertex(38,146){2.0}
\Line(38,146)(68,146)
\ArrowLine(92,130)(38,130)
\ArrowLine(38,130)(12,130)\Text(18,122)[]{$c$}
\Photon(38,146)(38,130){2}{3}\Text(54,138)[]{$W$}
\ArrowLine(68,146)(130,146) \Text(125,156)[]{$d$}
\Text(128,138)[]{\scriptsize{2}}
\Vertex(38,130){2.0}
\ArrowLine(92,80)(92,130)\Text(84,90)[]{$s$}
\ArrowLine(130,130)(100,130)\Text(125,120)[]{$b$}
\ArrowLine(100,130)(100,80)\Text(110,90)[]{$b$}
 \Text(97,84)[]{\scriptsize{3}}
\Text(74,65)[]{\bf (a)}
\Text(143,142)[]{$+$}
\ArrowLine(153,146)(199,146) \Text(163,156)[]{$u$}
\Text(155,138)[]{\scriptsize{1}}
\Vertex(178,146){2.0}
\Line(199,146)(229,146)
\Vertex(210,105){2.0}
\ArrowLine(219,146)(265,146) \Text(257,156)[]{$d$}
\Text(263,138)[]{\scriptsize{2}}
\Photon(178,146)(210,105){2}{6}\Text(198,138)[]{$W$}
\ArrowLine(210,130)(153,130)\Text(163,120)[]{$c$}
\ArrowLine(265,130)(218,130)\Text(257,120)[]{$b$}
\ArrowLine(210,80)(210,130)\Text(202,90)[]{$s$}
\ArrowLine(218,130)(218,80)\Text(226,90)[]{$b$}
 \Text(215,84)[]{\scriptsize{3}}
\Text(214,65)[]{\bf (b)}
\Text(279,142)[]{$+$}
\ArrowLine(288,146)(316,146) \Text(298,156)[]{$u$}
\Text(290,138)[]{\scriptsize{1}}
\Line(316,146)(360,146) 
\Vertex(316,130){2.0}
\Line(360,146)(390,146)
\Vertex(394,146){2.0}
\ArrowLine(390,146)(420,146) \Text(416,156)[]{$d$}
\Text(412,138)[]{\scriptsize{2}}
\PhotonArc(360,110)(50,45,155){2}{10}\Text(346,172)[]{$W$}
\ArrowLine(328,130)(288,130)\Text(298,120)[]{$c$}
\ArrowLine(328,80)(328,130)\Text(320,90)[]{$s$}
\ArrowLine(336,130)(336,80)\Text(346,90)[]{$b$}
\ArrowLine(420,130)(336,130)\Text(416,120)[]{$b$}
 \Text(333,84)[]{\scriptsize{3}}
\Text(354,65)[]{\bf (c)}
\end{picture}
\end{center}
\vspace{-2.2in}
\caption{W-exchange across quark lines.}
\end{figure}

\begin{figure}
\begin{center}\begin{picture}(600,100)(-150,40)
\ArrowLine(8,146)(38,146) \Text(18,156)[]{$s$}
\Text(10,138)[]{\scriptsize{1}}
\CArc(38,138)(8,-90,90)
\CArc(74,138)(8,90,270)
\Vertex(46,138){2.0}
\ArrowLine(38,130)(8,130)\Text(18,122)[]{$u$}
\Photon(46,138)(66,138){2}{4}\Text(56,148)[]{$W$}
\ArrowLine(74,146)(140,146) \Text(135,156)[]{$d$}
\Text(138,138)[]{\scriptsize{2}}
\Vertex(66,138){2.0}
\ArrowLine(92,130)(74,130)
\ArrowLine(92,80)(92,130)\Text(84,90)[]{$c$}
\ArrowLine(140,130)(100,130)\Text(135,120)[]{$b$}
\ArrowLine(100,130)(100,80)
\Text(97,84)[]{\scriptsize{3}}
\Text(110,90)[]{$b$}
\end{picture}
\end{center}
\vspace{-.7in}
\caption{Annihilation diagram.}
\end{figure}

\end{document}